\documentclass[aps,twocolumn,prl,showpacs,preprintnumbers,amsmath,amssymb]{revtex4}
\newcommand{\lsim} 
 {\ \raise.35ex\hbox{$<$}\kern-0.75em\lower.5ex\hbox{$\sim$}\ }
\newcommand{\gsim}
 {\ \raise.35ex\hbox{$>$}\kern-0.75em\lower.5ex\hbox{$\sim$}\ }

\usepackage{graphicx}
\usepackage{dcolumn}
\usepackage{bm}
\usepackage{amsmath}
\usepackage{times}
\usepackage[usenames]{color}
\usepackage{natbib}
\usepackage{ulem}

\begin{document}
\title{Quantum Melting of Magnetic Order in an Organic Dimer-Mott Insulating System} 
\author{Makoto Naka and Sumio Ishihara}
\affiliation{Department of Physics, Tohoku University, Sendai 980-8578, Japan}
\date{\today}
\begin{abstract}  
Quantum entanglement effects between the electronic spin and charge degrees of freedom are examined in an organic molecular solid, termed a dimer-Mott insulating system, in which molecular dimers are arranged in a crystal as fundamental units. 
A low energy effective model includes an antisymmetric exchange interaction, as one of the dominant magnetic interactions. 
This interaction favors a 90 degree spin configuration, and competes with the Heisenberg-type exchange interaction. Stabilities of the magnetic ordered phases are examined by using the spin-wave theory, as well as the  Schwinger-boson theory. 
It is found that the spin-charge interaction promotes an instability of the long-range magnetic ordered state around a parameter region where two spin-spiral phases are merged. 
Implication for the quantum spin liquid state observed in $\kappa$-(BEDT-TTF)$_2$Cu$_2$(CN)$_3$ is discussed. 
\end{abstract}

\pacs{75.10.Kt, 75.25.Dk, 75.30.Ds, 75.30.Et}

\maketitle
\narrowtext



%
%

%




%
\begin{figure}[t]
\begin{center}
\includegraphics[width=1.0\columnwidth, clip]{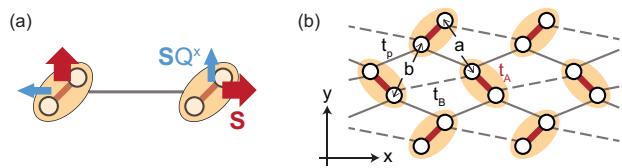}
\end{center}
\caption{(Color online) 
(a) A 90 degree spin configuration induced by the spin-charge coupling interaction. 
The thick and thin arrows represent directions of the spin moment (${\bm S}$) and the spin-charge moment (${\bm S}Q^{x}$), respectively. 
The circles and ellipses represent the BEDT-TTF molecules and dimers, respectively. 
(b) A schematic lattice structure and the dominant hopping integrals for $\kappa$-(BEDT-TTF)$_2$X in units of molecules. 
Inequivalent molecules are denoted by $a$ and $b$.
}
\label{fig:90deg}
\end{figure}

Quantum entanglements between electronic spin and other degrees of freedom in solids give rise to a number of exotic phenomena in correlated electron materials. 
Prototypical examples are seen in a strong coupling between the magnetic and electric polarizations in multiferroic materials~\cite{Kimura, Katsura}, and in a spin-charge coupling in magnetoresistive materials~\cite{Chahara, Helmolt}. 
They do not only induce new cross-correlated phenomena, but also trigger reexaminations of magnetic phenomena, which has been examined so far in proper magnets without other degrees of freedom~\cite{Zaanen, Zhang, Maekawa, Mila}. 
Most of the target materials have been searched in transition-metal compounds where robust spin polarizations emerge due to the strong electron correlation. 

Another class of the candidate materials, in which quantum entanglements between spin and other degrees of freedom are expected, is the organic molecular solids. 
Instead of the atomic $d$- and $f$-orbitals in transition-metal ions, magnetism is responsible for the molecular orbitals (MO). 
Although the electron-electron interaction is smaller than that in the transition-metal ions, 
some series of organic molecular solids are identified as strong correlated magnets due to the small overlap integral between MOs, and show a rich variety of exotic phenomena~\cite{Fukuyama}.

A dimer-Mott (DM) insulating system is one of the examples.
A paired molecule is a unit of a crystal lattice, and two outermost MOs in each dimer unit build the bonding and antibonding orbitals. 
When one hole/electron per dimer occupies these MOs in most of the materials, these are the Mott insulators in a case of strong electron-electron interaction~\cite{Kino}.  
Low dimensional organic solids, $\kappa$- and $\beta'$-(BEDT-TTF)$_2$X, and (TMTTF)$_2$Y  (X and Y are anion molecules), are known as example materials. 
Several exotic magnetic phenomena, such as quantum spin liquid state, superconductivity related to magnetic fluctuation, have been found in these materials~\cite{Kanoda, Miyagawa, Williams, Shimizu, sYamashita, mYamashita}. 
In addition to the spin degree of freedom, the electronic charge degree of freedom inside of the molecular dimer is recently enlightened; 
the dielectric anomalies are experimentally observed~\cite{Jawad, Lang, Iguchi, Dressel, Yakushi}, and are attributable to the local electric dipole moments inside of the dimer units~\cite{Naka, Hotta, Gomi, Clay, Naka2}. 
It is now the stage where the magnetic phenomena observed in the DM insulating materials should be reexamined by taking the spin-charge entanglement into account.  

In this Letter, we examine entanglement effects between the spin and charge degrees of freedom in a DM insulating system. 
A model Hamiltonian for the $\kappa$-(BEDT-TTF) type crystal lattice includes a novel spin-charge coupling, which gives rise to an antisymmetric exchange interaction. 
This interaction favors the 90 degree spin configuration between the nearest-neighboring (NN) BEDT-TTF dimers as shown in Fig.~\ref{fig:90deg}(a), and compete with the conventional Heisenberg exchange interaction. 
Mean field (MF) magnetic phase diagram and stabilities of the long-range magnetic ordered states are examined by the spin-wave (SW) approximation, as well as the Schwinger-boson (SB) MF approximation. 
It is found that this spin-charge coupling promotes an instability of the long-range magnetic ordered state around a parameter space, in which two spin-spiral phases are merged. 
This is attributed to the low-energy spin fluctuation induced through the spin-charge coupling. 
Implications of the present results  for the spin liquid state observed in $\kappa$-(BEDT-TTF)$_2$Cu$_2$(CN)$_3$ are discussed. 

We start from a tight-binding model Hamiltonian for the DM insulating system: the extended Hubbard model, where the BEDT-TTF molecules are identified as the fundamental units. 
This is given by 
\begin{align}
{\cal H}_{\rm exH}&= t_A\sum_{i \sigma} \left ( c_{i a \sigma}^\dagger c_{i b \sigma} + {\rm H.c.} \right )
+U\sum_{i \mu(=a,b)} n_{i \mu \uparrow} n_{i \mu \downarrow} \nonumber \\
&+V_{A} \sum_{i } n_{ia}n_{ib}+\sum_{\langle ij \rangle \sigma} t_{ij}^{\mu \mu'} 
\left ( c_{i \mu \sigma}^\dagger c_{j \mu' \sigma}+{\rm H.c.} \right)       \nonumber \\
&+\sum_{\langle ij \rangle \mu \mu'} V_{ij}^{\mu \mu'} n_{i \mu} n_{j \mu'} ,  
\label{eq:exh}
\end{align}
where 
$c_{i \mu \sigma}$ is an annihilation operator for a hole at the $\mu(=a, b)$ molecule in the $i$-th dimer unit with spin $\sigma(=\uparrow, \downarrow)$, 
and 
$n_{i \mu }=\sum_{\sigma}n_{i \mu \sigma}=\sum_{\sigma}c_{i \mu \sigma}^\dagger c_{i \mu \sigma}$ is the number operator. 
The first three terms represent the interactions inside of a dimer unit, i.e. 
the intra-dimer carrier hopping ($t_A$), the intra-molecular Coulomb interaction ($U$), and the inter-molecular Coulomb interaction inside of a dimer ($V_A$). 
The last two terms represent the interactions between the dimer units, i.e.  
the carrier hopping between the molecule $\mu$ in the $i$-th dimer and the molecule $\mu'$ in the $j$-th dimer ($t_{ij}^{\mu \mu'}$), and the Coulomb interactions ($V_{ij}^{\mu \mu'}$).  

The effective spin model derived from the extended Hubbard model introduced above is more useful to examine the low-energy magnetic states. 
When the intra-dimer interactions are sufficiently larger than the inter-dimer interactions, the number of holes in a dimer unit is restricted to be one, and the inter-dimer parts are treated as the perturbational interactions. 
The spin and charge degrees of freedom, respectively, inside of the dimer unit are described by the spin operators with an amplitude of 1/2 defined as ${\bm S} = (1/2) \sum_{ s s' \nu} { c}^{\dagger}_{i  \nu s} {\bm \sigma}_{s s'}  { c}_{i  \nu s'}$ and the pseudo-spin operator ${\bm Q} =(1/2) \sum_{s \nu \nu'} {\hat c}^{\dagger}_{i  \nu s } {\bm \sigma}_{\nu \nu'}  {\hat c}_{i \nu' s}$, where ${\bm \sigma}$ are the Pauli matrices, and ${\hat c}_{\nu s} = \sum_{\mu = a,b} W_{\nu \mu} c_{\mu s}$ 
with a unitary matrix $W =(\sigma^{z} + \sigma^{x})/\sqrt{2}$. 
The eigenstates for $Q_i^z$ with the eigenvalue of $1/2$ and $-1/2$, respectively, represent the states, in which one hole occupies the antibonding orbital $\psi_{\rm AB}=(\psi_a -\psi_b)/\sqrt{2}$ and the bonding orbital $\psi_{\rm B} =(\psi_a+\psi_b)/\sqrt{2}$, 
and those for $Q_i^x$ with $1/2$ and $-1/2$, respectively, represent the states, in which one hole occupies the $a$ orbital $(\psi_a)$, and the $b$ orbital $(\psi_b)$ inside of the dimer. 

Up to the second order perturbations, the Kugel-Khomskii type Hamiltonian~\cite{Khomskii} for the spin and charge degrees of freedom is obtained~\cite{Naka, Hotta}. 
This is expressed as a sum of a number of the exchange terms classified by the perturbation processes, and 
the full expression of the Hamiltonian is presented in Refs.~\cite{Naka, Naka2}. 
When the magnetic properties in the DM phase without the static charge polarization are focused on, we have confirmed that the following Hamiltonian plays as a minimal model which includes the dominant terms instead of all exchange terms: 
\begin{align} 
{\cal H} &= \sum_{\langle ij \rangle} J_{ij} {\bm S}_{i} \cdot {\bm S}_{j} -\Gamma \sum_{i} Q_{j}^{z} \notag \\
&+ \sum_{\langle ij \rangle} K_{ij} (Q_{i}^{x} - Q_{j}^{x}) {\bm S}_{i} \cdot {\bm S}_{j}. 
\label{eq:hamil}
\end{align}
The first and second terms in Eq.~(\ref{eq:hamil}), respectively, represent the Heisenberg interaction, and the intra-dimer carrier hopping, corresponding to the energy difference between the bonding and antibonding orbitals. 
The third term is the main term, i.e. the spin-charge coupling term.  
We have checked numerically that the magnetic phase diagram and magnetic order parameters in the DM phase calculated by this Hamiltonian reproduce qualitatively those by the Hamiltonian including all interaction terms, such as 
$\sum_{\langle ij \rangle} \tilde{J}_{ij}{\bm S}_i\cdot {\bm S}_j Q^x_i Q^x_j$ 
and $\sum_{\langle ij \rangle}W_{ij}Q^x_i Q^x_j$. 
Details are presented in the Supplemental Material (SM). 

Let us first consider the spin-charge coupling on an isolated bond connecting NN dimers in the $\kappa$-(BEDT-TTF) type lattice [Fig.~\ref{fig:90deg}(b)].
It is noticeable that this spin-charge coupling term provides an antisymmetric interaction; its sign is changed by interchanging sites $i$ and $j$. 
Any rotations of the local pseudo-spin frames do not remove this alternation of signs, when we express this term for all NN bonds in a unified fashion. 
In the DM phase, where the local charge polarizations and currents are zero, i.e. $\langle Q^x \rangle=\langle Q^y \rangle=0$, 
the single-site wave function is given by a form 
$\left| \psi \right\rangle = a \left|\uparrow, \uparrow \right\rangle + b \left|\downarrow, \downarrow \right\rangle$, with complex numbers $a$ and $b$. 
The bracket represents $\left|S^{z}, Q^{z} \right\rangle$ where $S^{z}$ is taken as the local spin quantization axis.
That is, the spin and charge sectors are entangled strongly with each other. 
The spin and spin-charge coupled moments in this wave function are calculated as  
$\langle {\bm S} \rangle = (0, 0,  (|a|^{2} - |b|^{2})/2)$ and $\langle {\bm S} Q^{x} \rangle = ({\rm Re} [a^{*}b]/2,  {\rm Im} [a^{*}b]/2, 0)$, respectively. 
Thus, $\langle {\bm S} \rangle$ is perpendicular to $\langle {\bm S} Q^{x} \rangle$. 
As shown in Eq.~(\ref{eq:hamil}), the spin-charge coupling term is given by inner products of $ {\bm S}_i Q^{x}_{i}$ and ${\bm S}_j $, and ${\bm S}_j Q^{x}_{j}$  and $ {\bm S}_i $, 
with the positive and negative coupling constants, respectively. 
Therefore,  $ {\bm S}_i Q^{x}_{i}$  and $ {\bm S}_j $ ($ {\bm S}_j Q^{x}_{j}$  and $ {\bm S}_i $), tend to be antiparallel (parallel) with each other. 
Consequently, the effective interaction between ${\bm S}_i$ and ${\bm S}_j$ favors a spin configuration with 90 degree as shown in Fig.~\ref{fig:90deg}(a), in a similar way to the Dzyaloshinsky-Moriya interaction. 

\begin{figure}[t]
\begin{center}
\includegraphics[width=1.0\columnwidth, clip]{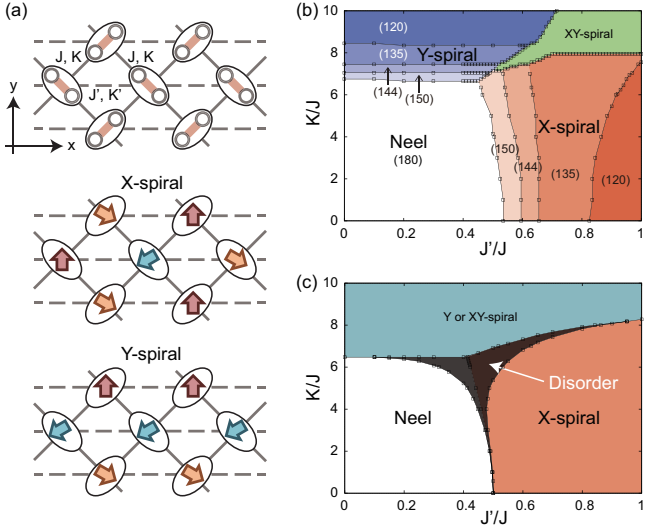}
\end{center}
\caption{
(Color online) 
(a) A schematic lattice structure and the exchange interactions for $\kappa$-(BEDT-TTF)$_2$X in units of dimers (upper panel). 
Schematic spin structures in the X- and Y-spiral phases are shown in the middle and lower panels, respectively. 
The circles and ellipses represent the BEDT-TTF molecules and dimers, respectively. 
(b) Ground state magnetic phase diagram obtained by the MF approximation. 
Numbers in the parentheses represent angles between the NN spins along the $x$ and $y$ axes in the X- and Y-spiral phases, respectively, as a unit of degree. 
(c) Ground state magnetic phase diagram in which fluctuation effects are taken into account by the linear SW approximation. 
An infinite number sites along the $x$ axis is adopted. 
The shaded area denoted by ``Disorder" represents a parameter space in which the magnetic moment obtained by the linear SW approximation is negative.　
}
\label{fig:pd}
\end{figure}

In the $\kappa$-(BEDT-TTF) type lattice, since the number of the NN bonds along the $y$ axis is larger than that along the $x$ axis, the spin-charge coupling term favors a 90 degree spin-spiral order along the $y$ axis in the DM phase. 
In a realistic parameter set, a magnitude of the spin-charge coupling is comparable to the Heisenberg-type interaction given by the first term in Eq.~(\ref{eq:hamil}). 
We will show that the competition between these two interactions induces instability of the conventional spin order and promotes a spin disordered state. 
As shown in Fig.~\ref{fig:pd}(a), there are two kinds of the exchange interactions: 
the interactions between the equivalent dimers and the interactions between the inequivalent dimers denoted by $(J', K')$ and $(J, K)$, respectively. 
In the following analyses, for simplicity, we adopt $J$ as a unit of energy, and a relation $J'/J=K'/K$ and $\Gamma=40J$ are assumed. 

Spin structures under the competition between the interactions in Eq.~(\ref{eq:hamil}) is examined by the linear SW approximation based on the MF approximation. 
As the MFs, we introduce $\langle S^{\mu} \rangle $, $\langle Q^{\nu} \rangle$, and $\langle S^{\mu} Q^{\nu} \rangle$ where $(\mu, \nu) = (x, y, z)$.
The MF solutions are calculated in finite size clusters up to the $6 \times 6$ unit cells under the periodic boundary condition.
Three kinds of the Holstein-Primakoff (HP) bosons are introduced; the spin excitation, charge excitation, and spin-charge coupled excitation.  
We introduce the local unitary transformations ${\cal U}$, which turns the local quantization axes to be parallel to the MF spin and pseudo-spin directions at each site, 
as $\widetilde{S}^{z} = {\cal U} S^{z} {\cal U}^{\dagger}$ and $\widetilde{Q}^{z} = {\cal U} Q^{z} {\cal U}^{\dagger}$. 
When the local spin and charge states are represented by $\left|\widetilde{S}^{z}, \widetilde{Q}^{z}\right\rangle$, 
$\left| \uparrow, \uparrow \right\rangle( \equiv \left| 1 \right\rangle)$ is the MF ground state, 
and 
$\left| \downarrow, \uparrow \right\rangle (\equiv \left| 2 \right\rangle)$, 
$\left| \uparrow, \downarrow \right\rangle (\equiv \left| 3 \right\rangle)$, and 
$\left| \downarrow, \downarrow \right\rangle (\equiv \left| 4 \right\rangle)$ are the excited states. 
The excitations are given by the generators $J^{n}_{m}$ in the SU(4) algebra as 
$J^{n}_{m} \left| n \right \rangle =\left| m \right \rangle$ where $m$ and $n$ take $1$--$4$. 
The HP bosons are introduced by the transformations $J^{1}_{1} = M- \sum_{n \neq 1} \alpha^{1 \dagger}_{n} \alpha^{1}_{n}$, $J^{1}_{n} = \alpha^{1 \dagger}_{n} \sqrt{M - \sum_{l \neq 1} \alpha^{1 \dagger}_{l} \alpha^{1}_{l}}$, and $J^{l}_{n} = \alpha^{1 \dagger}_{n} \alpha^{1}_{l}$ 
for $(l, n) \neq 1$ with $M=1$. 
The Hamiltonian in Eq.~(\ref{eq:hamil}) is rewritten by the HP bosons within the linear SW approximation where $1/M$ expansion is applied. Details are given in SM. 

%
A MF magnetic phase diagram at zero temperature is presented in a plane of the anisotropy in the exchange interaction ($J'$) and the spin-charge coupling constant ($K$) in Fig.~\ref{fig:pd}(b). 
The horizontal axis at $K=0$ is identical to the Heisenberg model. 
The conventional N$\rm \acute {e}$el order and the coplanar 120 degree structure, respectively, are realized at $J'=0$ and $1$, which correspond to the square and equilateral triangular lattices. 
Between them, a spiral spin structure characterized by a momentum $(q_x, 0)$, termed a ``X-spiral", appears. 
When the spin-charge coupling $K$ turns on, a different spiral spin structure characterized by $(0, q_y)$, termed a ``Y-spiral", emerges in the small $J'$ region. 
Spin structures in the X- and Y-spiral phases are shown in Fig.~\ref{fig:pd}(a). 
The Y-spiral phase is consequence of the competition between the Heisenberg interaction and the spin-charge coupling which favors the 90 degree spin configuration as mentioned above. 
The X-spiral, Y-spiral, and N$\rm \acute {e}$el ordered phases are merged around $(J'=0.45, K=6.5)\equiv (J'_c, K_c)$. 
Another spiral phase termed ``XY-spiral" also appears between the X- and Y-spiral phases, where the spin structures are characterized by both the $x$ and $y$ components of the momentum. 

A possibility of the spin disorder phase is examined by calculating the ordered magnetic moment corrected by the quantum fluctuations in the linear SW approximation. 
This is given by $m = M/2 - N^{-1} \sum_{\bm k} ( \langle \alpha^{1 \dagger}_{2}({\bm k}) \alpha^{1}_{\rm 2}({\bm k}) \rangle + \langle \alpha^{1 \dagger}_{\rm 4}({\bm k}) \alpha^{1}_{4}({\bm k}) \rangle )$, where the second and third terms are the quantum corrections due to the spin and spin-charge coupled excitations, respectively. 
\begin{figure}[t]
\begin{center}
\includegraphics[width=1.0\columnwidth, clip]{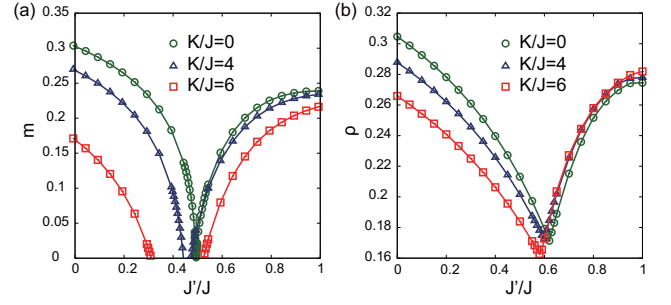}
\end{center}
\caption{(Color online) 
(a) Magnetic moments calculated by the SW method as functions of the exchange anisotropy ($J'/J$) for several values of the spin-charge coupling. 
(b) Condensation densities calculated by the SB MF approximation. 
}
\label{fig:mz}
\end{figure}
%
The results are plotted as a function of $J'$ in Fig.~\ref{fig:mz}(a) for several values of $K$. 
In the case of no spin-charge coupling ($K=0$), $m$ decreases toward the phase boundary ($J'=0.5$) between the commensurate (C) N$\rm \acute{e}$el ordered phase and the incommensurate (IC) X-spiral phase, i.e. the C-IC transition point as known in Ref.~\cite{Trumper}. 
This reduction becomes pronounce with increasing the spin-charge coupling, and $m$ is negative in a finite region of $J'$. 
We survey any possible MF magnetic and charge ordered states described by $\langle S^\mu_i \rangle$, $\langle Q_i \rangle$, and $\langle S_i^\mu Q_i^\nu \rangle$ up to the $6 \times 6$ site clusters. 
A part of the calculations are performed in a cluster with an infinite number of sites along the $x$ axis. 
The above results imply that, at least, assumed magnetic orders are unstable, although the stable magnetic structure in the parameter range, in which $m$ is negative, can not be identified in the present calculations. 
The phase diagram where the quantum fluctuation is taken into account in the linear SW approximation is presented in Fig.~\ref{fig:pd}(c). 
The parameter region where $m$ is negative is represented by a shaded area, 
and appears around the point at which the two C-IC transitions are merged with each other. 
That is, the competition between the Heisenberg interaction and the spin-charge coupling enhances the instability of the magnetic ordered phase. 
\begin{figure}[t]
\begin{center}
\includegraphics[width=1.0\columnwidth, clip]{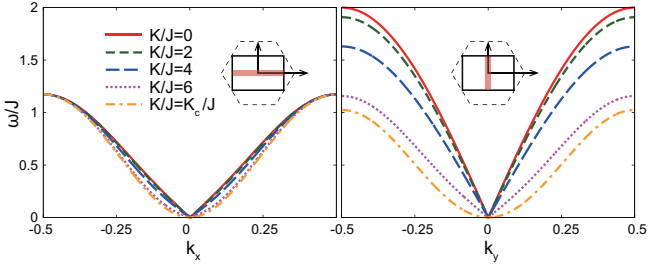}
\end{center}
\caption{(Color online) 
Dispersion relations of SW along the $k_{x}$ (left panel) and $k_{y}$ (right panel) axes for several values of the spin-charge coupling. 
A parameter value is chosen to be $J'/J = J'_{\rm c}/J$. 
}
\label{fig:ek}
\end{figure}

The results introduced above are checked by utilizing the SB MF theory~\cite{Arovas, Hirsch, Gazza, Mezio}. 
We introduce the boson operator, $\beta_{i \sigma \tau}^\dagger$ where $\sigma(=\uparrow, \downarrow)$ and $\tau(=\uparrow, \downarrow)$ are the spin and charge subscripts, respectively. 
These operators produce the eigenstates of $S^{z}$ and $Q^{z}$ as 
$\left|\sigma \tau \right\rangle = \beta_{i \sigma \tau}^\dagger \left | 0 \right\rangle$, where $\left| 0 \right\rangle$ is the vacuum for the SBs. 
The Hamiltonian in Eq.~(\ref{eq:hamil}) is rewritten by the SB scheme as 
\begin{align}
&{\cal H} = -\Gamma \sum_{i} Q_{i}^{z} + \sum_{\langle ij \rangle \tau \tau'} J_{ij} \left[{\cal N}(B^{\tau \tau' \dagger}_{ij} B^{\tau \tau'}_{ij}) - A^{\tau \tau' \dagger}_{ij} A^{\tau \tau'}_{ij} \right] \notag \\
&+ \sum_{\langle ij \rangle \tau \tau'} \frac{1}{2} C_{ij}^\tau C_{ij}^{\tau'} K_{ij} \left[{\cal N}(B^{\tau {\bar \tau} \dagger}_{ij} B^{\tau' \tau'}_{ij}) - A^{\tau {\bar \tau} \dagger}_{ij} A^{\tau' \tau'}_{ij} + {\rm H.c.} \right],   
\label{eq:sb}
\end{align}
where 
$A^{\tau \tau'}_{ij} = \frac{1}{2} (\beta_{i \uparrow \tau} \beta_{j \downarrow \tau'} - \beta_{i \downarrow \tau} \beta_{j \uparrow \tau'})$ 
and 
$B^{\tau \tau'}_{ij} = \frac{1}{2} (\beta^{\dagger}_{i \uparrow \tau} \beta_{j \uparrow \tau'} + \beta^{\dagger}_{i \downarrow \tau} \beta_{j \downarrow \tau'})$. 
We define the normal product $\cal N$, and  ${\bar \tau}=\uparrow (\downarrow)$ for $\tau=\downarrow (\uparrow)$. 
A numerical factor takes $C_{ij}^{\tau} = 1$ on the horizontal bonds along the $x$ axis, 
and $C_{ij}^{\tau} =1$ $(-1)$ for $\tau =\uparrow$ $ ( \downarrow)$ on the diagonal bonds. 
The MF-type  decoupling is introduced in the bilinear terms in Eq.~(\ref{eq:sb}) as  
$X_{ij} Y_{ij} \approx \langle X_{ij} \rangle Y_{ij} + X_{ij} \langle Y_{ij} \rangle - \langle X_{ij} \rangle \langle Y_{ij} \rangle$ 
where $X_{ij}$ and $Y_{ij}$ are the bond operators, 
and the expectation values are determined selfconsistently under the global constraint $\sum_{i \sigma \tau} \langle \beta^{\dagger}_{i \sigma \tau} \beta_{i \sigma \tau} \rangle = 2N$ ($N$ is the number of unit cells). 
In order to examine a possibility of the instability of long-range ordered state, 
we calculate the SB condensation densities $(\rho)$ corresponding to the sublattice magnetization. 
This is expressed by the coefficients in the Bogoliubov transformation, which diagonalizes  the SB MF Hamiltonian.  
Details are presented in SM. 
The calculated condensation densities are shown in Fig.~\ref{fig:mz}(b). 
In the case of $K=0$, a reduction of $\rho$ is seen around $J'=0.6$ corresponding to the C-IC transition point. 
Similar to the case of the SW approximation, the reduction is increased with increasing the spin-charge coupling. 

Through the calculations by the two different approximation methods, 
we believe that an instability of the magnetic ordered state around the ``double" C-IC transition point is not an artifact, but is robust. 
We note that $\rho$ does not reach zero even in its minimum points, and the assumed spin structure is still stable within this SB MF approximation. 
It is well known that the linear SW approximation overestimates the magnetic disordered state, while the SB MF approximation underestimate it. 
Thus, we expect that the real situation for the reduction of $m$ and emergence of the quantum disordered state are in between the two results.  
To elucidate the low-energy spin fluctuation causing the reduction of  the magnetic moment, we show the variations of the SW dispersions for the change of the spin-charge coupling. 
The energy dispersions for the SW excitations along the $k_x$ and $k_y$ axes in the N$\rm \acute{e}$el ordered phase near the C-IC transition points are shown in Figs.~\ref{fig:ek}(a) and (b), respectively. 
When the system approaches to the double C-IC point, the linear dispersion around the $\Gamma$ point is changed into the quadratic-like dispersion in both the $k_x$ and $k_y$ axes. 
Simultaneously, the bandwidth along the $k_y$ direction becomes narrow. 
These changes lead to large enhancement of the low energy spin fluctuation, and the reduction of the ordered magnetic moment. 

Finally, we touch briefly implications of the present results for the experimentally observed quantum spin liquid state in $\kappa$-(BEDT-TTF)$_2$Cu$_2$(CN)$_3$. 
At first, the present theory is based on the assumption that the static charge polarization inside of the dimers does not occur and the system is in the DM insulating state. 
This is consistent with the experimental results in $\kappa$-(BEDT-TTF)$_2$Cu$_2$(CN)$_3$ where a clear charge ordering transition has not been observed~\cite{Dressel, Yakushi}. 
According to the first-principles calculation~\cite{Nakamura}, 
the anisotropy in the electron hopping integrals is estimated to be $t'/t \approx 0.8$ corresponding to $J'/J \approx 0.64$. 
That is, this material is located near the C-IC transition point at $J'/J \approx 0.5$ rather than the isotropic point $J'/J = 1$. 
A value of the spin-charge coupling constant is also estimated to be $K/J \approx 2$--$5$. 
Thus, in the magnetic phase diagram shown in Fig.~\ref{fig:pd}(c), 
$\kappa$-(BEDT-TTF)$_2$Cu$_2$(CN)$_3$ is located in or close to the shaded area in which the long-range magnetic ordered state is not stable.  

The authors would like to thank J. Nasu and T. Watanabe for valuable discussions. 
M.N. also acknowledges D. Yamamoto for fruitful discussions. 
This work is supported by JST CREST, and Grant-in-Aid for Scientific Research (No. 26287070 and No. 15H02100) from MEXT (Japan). 
A part of the numerical calculations has been performed using the supercomputing facilities at ISSP, the University of Tokyo. 


%
%
%
%
%
\clearpage
%
\section*{\Large 
Supplemental Material for the article \\``Quantum Melting of Magnetic Order in an Organic Dimer-Mott Insulating System"
}
In this Supplemental Material, detailed calculations for the effective model Hamiltonian, the spin-wave approximation, and the Schwinger boson mean-field approximation are presented. 
\section{Effective model Hamiltonian} 
In this section, we present validity of the effective model Hamiltonian given in Eq.~(2) in the main text (MT).  
A full expression of the effective model Hamiltonian derived from the extended Hubbard model is presented in Refs.~\cite{Naka_si, Naka2_si}. 
This is divided into the following three terms; 
\begin{align}
{\cal H}_{\rm eff} = { {\cal H}}_{\rm intra} + { {\cal H}}_{V} + {\cal H}_{\rm ex} .
\end{align}
The first two terms are represented by 
\begin{align}
{ {\cal H}}_{\rm intra} + { {\cal H}}_{V} = - \Gamma \sum_{i} Q^{z}_{i} + \sum_{\langle ij \rangle} W_{ij} Q^{x}_{i} Q^{x}_{j}, 
\label{eq:intra2}
\end{align}
where $\Gamma$($=2t_{A}$) is the intra-dimer electron hopping and $W_{ij}$($=V^{aa}_{ij} + V^{bb}_{ij} - V^{ab}_{ij} - V^{ba}_{ij} $) is the inter-dimer Coulomb interaction.  
The third term represents the exchange interactions originating from the second order perturbation with respect to the inter-dimer hopping and is classified by the spin-singlet and spin-triplet intermediate states in the perturbational  processes as 
\begin{align}
{\cal H}_{\rm ex} = - \sum_{\langle ij \rangle} \left( \frac{3}{4} + {\bm S}_{i} \cdot {\bm S}_{j} \right) h^{T}_{ij} - \sum_{\langle ij \rangle} \left( \frac{1}{4} - {\bm S}_{i} \cdot {\bm S}_{j} \right) h^{S}_{ij} .  
\label{eq:ex}
\end{align}
Here, $h^T_{ij}$ and $h_{ij}^S$ are represented by the pseudo-spin operators at sites $i$ and $j$, as shown in Refs.~\cite{Naka_si,Naka2_si}. 

\begin{figure}[t]
\begin{center}
\includegraphics[width=1.0\columnwidth, clip]{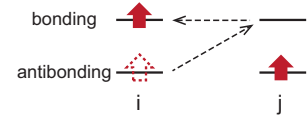}
\end{center}
\caption{(Color online) 
One of the dominant perturbational processes between the neighboring $i$- and $j$-th dimers contributing to the spin-charge coupling term in Eq.~(\ref{eq:sc1}). }
\label{fig:proc}
\end{figure}

The spin-charge coupling term of the present interest is given by
\begin{align}
{\cal H}_{\rm ex}^K=\sum_{\langle ij \rangle} 
K_{ij} (Q_{i}^{x} - Q_{j}^{x}) {\bm S}_{i} \cdot {\bm S}_{j}  .
\label{eq:sc1}
\end{align}
The dominant perturbational process in this term is shown in Fig.~\ref{fig:proc}, which consists of the cross terms of the diagonal and off-diagonal hopping integrals. 
In addition to this term,  
the following two terms are the dominant exchange interactions among several terms in ${\cal H}_J$:  the another type of the spin-charge coupling given by \begin{align}
{\cal H}_{\rm ex}^{\tilde J}= 
\sum_{\langle ij \rangle} {\tilde J}_{ij}{\bm S}_{i} \cdot {\bm S}_{j} Q^{x}_{i} Q^{x}_{j} ,
\label{eq:hei}
\end{align}
and the Heisenberg exchange interaction given by 
\begin{align}
{\cal H}_{\rm ex}^{J}=\sum_{\langle ij \rangle} J_{ij}{\bm S}_{i} \cdot {\bm S}_{j}  . 
\label{eq:sc2}
\end{align}
The exchange parameters $K_{ij}$, $J_{ij}$ and $\tilde J_{ij}$ are represented by the energy parameters of the extended Hubbard model given in Eq.~(1) in MT, and all of them are positive. 
Their magnitudes are the same orders with each other.  

\begin{figure}[t]
\begin{center}
\includegraphics[width=1.0\columnwidth, clip]{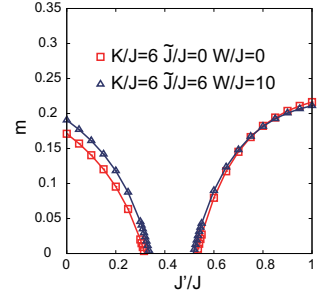}
\end{center}
\caption{(Color online) 
Magnetic moments calculated by the SW method. Square and triangle symbols, respectively, are obtained in the model Hamiltonian ${\cal H}$ adopted in MT, and ${\cal H}'$ where other dominant terms are added (see the text). 
}
\label{fig:mm}
\end{figure}

In MT, we focus on the DM phase where the local charge polarizations and currents are zero, i.e. $\langle Q^x \rangle=\langle Q^y \rangle=0$. 
We adopt the Hamiltonian ${\cal H}={\cal H}_{\rm intra}+{\cal H}_{\rm ex}^{J}+{\cal H}_{\rm ex}^{K}$ as a minimum model, and show that the spin-charge coupling promotes an instability of the long-range magnetic ordered phase. 
In order to check validity of this model, we calculate the magnetic moment by the linear spin wave (SW) method, where the remaining dominant terms are taken into account, i.e. ${\cal H}' \equiv {\cal H}+{\cal H}_{V}+{\cal H}_{\rm ex}^{\tilde J}$. 
The calculated results are shown in Fig.~\ref{fig:mm}, and are compared with the results obtained in ${\cal H}$ which is presented in Fig.~3(a) in MT. 
The magnetic moment obtained in ${\cal H}'$ shows negative values around $J'/J=0.4$, and difference between the two results are within ten percent.

%
%

\section{spin-wave approximation} 
\begin{figure}[t]
\begin{center}
\includegraphics[width=1.0\columnwidth, clip]{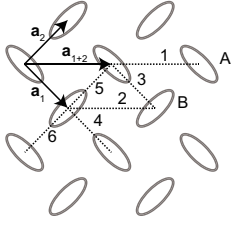}
\end{center}
\caption{(Color online) 
A schematic two dimensional plane in the $\kappa$-(BEDT-TTF)$_2$X crystal structure. 
Ovals denoted by A and B represent inequivalent dimers. 
Dotted lines and arrows are the inequivalent bonds and the primitive translation vectors, respectively. 
}
\label{fig:si}
\end{figure}

In this section, we present a detailed explanation for the linear SW approximation based on the mean-field (MF) method. 
As for the MF solutions, the spin moment $\langle {\bm S} \rangle$ and the spin-charge moment $\langle {\bm S}Q^{x} \rangle$, respectively, at each dimer unit are confined on the $S^{z}$--$S^{x}$ and $S^{z}Q^{x}$--$S^{x}Q^{x}$ planes, and 
the MF solutions are found by rotating them around the $S^{y}$ and $S^{y}Q^{x}$ axes. 
The unitary transformation for the rotating frame is given by ${\cal U} = {\cal U}_{S} {\cal U}_{SQ}$ with the unitary matrices ${\cal U}_{S} = \exp[- i \theta S^{y}]$ and ${\cal U}_{SQ} = \exp[- i \phi (2 S^{y} Q^{x})]$ where $\theta$ and $\phi$ are the site dependent angles. 
Explicitly, the $z$ components of the operators are represented in the rotating frames as 
\begin{align}
\widetilde{S}^{z} 
&
= \left ( S^{z} \cos\phi + 2S^{x}Q^{x} \sin\phi  \right ) \cos\theta  \nonumber \\
&+ \left ( S^{x} \cos\phi - 2S^{z}Q^{x} \sin\phi  \right ) \sin\theta, \\
\widetilde{Q}^{z} 
&
= Q^{z} \cos\phi - 2 S^{y}Q^{y}\sin\phi. 
\label{eq:uni}
\end{align}
In the local frame at each site, 
the spin and charge states are described by the wave function 
$\left| \widetilde{S}^{z}, \widetilde{Q}^{z} \right\rangle$. 
Thus, $\left| \uparrow, \uparrow \right\rangle (\equiv \left| 1 \right\rangle)$ 
is the MF ground state, and other three 
$\left| \downarrow, \uparrow \right\rangle (\equiv \left| 2 \right\rangle)$, $\left| \uparrow, \downarrow \right\rangle (\equiv \left| 3 \right\rangle)$, and $\left| \downarrow, \downarrow \right\rangle (\equiv \left| 4 \right\rangle)$ are the excited states. 
We introduce the generators in the SU(4) Lie algebra, $J_{m}^{l}(i) \ (l, m=1$--$4)$, which 
changes the spin and orbital states at site $i$ as $J_{m}^{l}(i) \left| n \right\rangle = \delta_{nl} \left| m \right\rangle$. 
The spin operators $\widetilde{S}^{\mu}_i$, the pseudo-spin operators $\widetilde{Q}^{\nu}_i$, and their products $\widetilde{S}^{\mu}_i \widetilde{Q}^{\nu}_i$ are represented by the linear combinations of $J_{m}^{l}(i)$.
For example, $\widetilde{S}^{z}_i = (1/2) \left [J_{1}^{1}(i) - J_{2}^{2}(i) + J_{3}^{3}(i) - J_{4}^{4}(i) \right ]$. 
Then, the generalized Holstein-Primacoff (HP) transformations in the SU(4) algebra are introduced as 
\begin{align}
J_{1}^{1}(i) 
&= M - \sum_{n} \alpha_{n}^{1\dagger}(i) \alpha_{n}^{1}(i),    \nonumber \\
J_{n}^{1}(i) 
&= \alpha_{n}^{1\dagger}(i) \sqrt{M - \sum_{l} \alpha_{l}^{1\dagger}(i) \alpha_{l}^{1}(i)}, 
 \nonumber \\
J_{n}^{l}(i) 
&= \alpha_{n}^{1\dagger}(i)\alpha_{l}^{1}(i), 
\label{eq:hp}
\end{align}
where both $n$ and $l$ do not take $1$, and $M=1$. 
The boson operators $\alpha_{n}^{1}$ describe the local spin, charge, and spin-charge excitations.  
Since the inequivalent two dimers exist in a unit cell of the $\kappa$-(BEDT-TTF)$_2$X crystal as shown in Fig.~\ref{fig:si}, we introduce two kinds of the HP bosons, $a_n^l(i)$ and $b_n^l(i)$, for the A and B sublattices, respectively, in each unit cell. 

%
%
The right-hand sides in Eqs.~(\ref{eq:hp}) are expanded up to the second order of the HP bosons. 
Then, the Hamiltonian presented in Eq.~(2) in MT is represented as the quadratic form of the boson operators as  
\begin{align}
{\cal H}_{\rm SW} = 
\frac{1}{2} \sum_{\bm k} {\bm \psi}^{\dagger}_{\bm k} {\cal D}_{\bm k} {\bm \psi}_{\bm k} + {\rm const.}, 
\end{align}
where ${\bm \psi}_{\bm k}$ is a set of the bosonic operators with the momentum $\bm k$ defined by  
\begin{align}
{\bm \psi}_{\bm k} &= 
\bigl[
a^1_2(\bm k), b^1_2(\bm k), a^1_3(\bm k), b^1_3(\bm k), a^1_4(\bm k), b^1_4(\bm k), 
\nonumber \\
& {a^1_2}^{\dagger}(-{\bm k}), {b^1_2}^{\dagger}(-{\bm k}), 
{a^1_3}^{\dagger}(-{\bm k}), {b^1_3}^{\dagger}(-{\bm k}), {a^1_4}^{\dagger}(-{\bm k}), {b^1_4}^{\dagger}(-{\bm k}) \bigr], 
\end{align} 
and ${\cal D}_{\bm k}$ is a 12$\times$12 matrix.
Through the Bogoliubov transformation, the Hamiltonian is diagonalized as  
\begin{align}
{\cal H}_{\rm SW} 
= \sum_{{\bm k} n=1}^{6} \epsilon_{{\bm k}n} \left (\gamma^{\dagger}_{{\bm k} n} \gamma_{{\bm k} n} + \frac{1}{2} \right) + {\rm const.}, 
\end{align}
with the energy $\epsilon_{{\bm k}n}$. A set of the bosonic operators are defined by the Bogoliubov transformation as ${\bm \gamma}_{\bm k} = {\cal T}_{\bm k} {\bm \psi}_{\bm k}$, which is explicitly represented as 
\begin{align}
{\bm \gamma}_{\bm k} 
&= \bigl( \gamma_{{\bm k}1}, \gamma_{{\bm k}2}, \gamma_{{\bm k}3}, \gamma_{{\bm k}4}, \gamma_{{\bm k}5}, \gamma_{{\bm k}6}, \nonumber \\
&  
\gamma^{\dagger}_{{-\bm k}1}, \gamma^{\dagger}_{{-\bm k}2}, \gamma^{\dagger}_{{-\bm k}3}, \gamma^{\dagger}_{{-\bm k}4}, \gamma^{\dagger}_{{-\bm k}5}, \gamma^{\dagger}_{{-\bm k}6} \bigr). 
\end{align}

Next, we define the magnetizations at sublattices A and B as 
\begin{align}
m_{\rm A(B)} = \frac{1}{N} \sum_{i \in {\rm A(B)}} \langle {\hat{S}^{z}_{i}} \rangle, 
\end{align}
where $\hat{S}^{z} = {\cal U}_{S} S^{z} {\cal U}^{\dagger}_{S}$.
This is rewritten in the linear SW scheme as 
\begin{align}
m_{\rm C} = \frac{M}{2} - \frac{1}{N} \sum_{{\bm k}}
\left[ \langle {\alpha^{1}_{2}}^{\dagger}({\bm k}) {\alpha^{1}_{2}}({\bm k}) \rangle 
+ \langle {\alpha^{1}_{4}}^{\dagger}({\bm k}) {\alpha^{1}_{4}}({\bm k}) \rangle \right], 
\end{align}
where $\alpha_n^l$ is $a_n^l$ and $b_n^l$ when ${\rm C}={\rm A}$ and ${\rm B}$, respectively. 
The 2nd and 3rd terms in the right-hand side represent the quantum corrections for the spin moment due to the spin and spin-charge fluctuations, respectively. 
In the N$\rm {\acute e}$el order state, we have the following results,  
\begin{align}
m_{\rm C} = \frac{M}{2} - \frac{1}{N} \sum_{\bm k} \sum_{n=7}^{12} \left[ \left| \left({\cal T}^{-1}_{\bm k} \right )_{l n} \right|^{2} + \left| \left ({\cal T}^{-1}_{\bm k} \right )_{m n} \right|^{2} \right], 
\end{align}
where $(l, m)=(1,5)$ and $(2,6)$ for the A and B sublattices, respectively. 
In the X-spiral phase introduced in MT,     
we obtain magnitude of the order parameter given as 
\begin{align}
m_{\rm C} 
&= \frac{M}{2} \cos \phi  \nonumber \\
&- \frac{1}{N} \sum_{\bm k} \sum_{n=7}^{12} 
\left[ \left| \left ({\cal T}^{-1}_{\bm k} \right)_{i n} \right|^{2} 
+ \left| \left ({\cal T}^{-1}_{\bm k} \right)_{j n} \right|^{2} \right] \cos \phi \notag \\ 
&- \frac{1}{N} \sum_{\bm k} \sum_{n=7}^{12} 
{\rm Re} \left[ \left({\cal T}^{-1}_{\bm k} \right)^{*}_{k n} 
\left({\cal T}^{-1}_{\bm k} \right)_{l n} \right] \sin \phi, 
\end{align}
where $(i, j, k, l)=(1, 5, 1, 3)$ and $(2, 6, 2, 4)$ for C=A and B, respectively, 
and a factor $\cos \phi$ in the first and second lines represent reduction of the local spin moment due to canting of the $\hat{S}^{z}$ axis toward the $S^{x}Q^{x}$ and $S^{z}Q^{x}$ axes, and the third term originates from the couplings between the spin and charge excitations. 
In the case where dominant parts of the electron hopping integrals given in Fig.~1(b) in MT are considered, 
the sublattice A and B are equivalent with each other and $m_{A} = m_{B}$ is satisfied. 
\section{Schwinger-boson mean-field approximation}
In this section, we introduce a detailed explanation for the Schwinger boson (SB) MF approximation adopted in MT. 
We generalize the conventional SB MF theory~\cite{Arovas_si, Hirsch_si, Gazza_si, Mezio_si}, which has been applied into the Heisenberg model, to the examination of the present spin-charge coupled model. 
We introduce the four boson operators $\beta_{i \sigma \tau}$ where 
$\sigma(=\uparrow,\downarrow)$ and $\tau(=\uparrow,\downarrow)$ are the subscripts for the spin and charge degrees of freedom, respectively, and $i$ represents a site.  
The eigenstates of $S^{z}$ and $Q^{z}$ are given by $\left| \sigma \tau \right\rangle = \beta^{\dagger}_{i \sigma\tau} \left| 0 \right\rangle$, where $|0\rangle$ is a vacuum of the boson. 
The following local constraint is required at each site: 
$\sum_{\sigma\tau} \beta^\dagger_{i \sigma\tau}\beta_{i \sigma\tau} = 1$. 
We introduce the bond operators defined by 
\begin{align} 
{A}^{\tau\tau'}_{ij} 
&= \frac{1}{2} (\beta_{i \uparrow \tau}\beta_{j \downarrow\tau'} - \beta_{i \downarrow \tau}\beta_{j \uparrow \tau'}), \notag \\
{B}^{\tau \tau'}_{ij} 
&= \frac{1}{2} (\beta^{\dagger}_{i \uparrow \tau}\beta_{j \uparrow \tau'} + \beta^{\dagger}_{i \downarrow \tau}\beta_{j \downarrow \tau'}) . 
\end{align}
Then, the Hamiltonian in Eq.~(2) in MT is rewritten as 
\begin{align}
{\cal H} &= - \Gamma \sum_{i} Q_{i}^{z} + \sum_{\langle ij \rangle \tau \tau'} J_{ij} \left[{\cal N}(B^{\tau \tau' \dagger}_{ij} B^{\tau \tau'}_{ij}) - A^{\tau \tau' \dagger}_{ij} A^{\tau \tau'}_{ij} \right] \notag \\
&+ \sum_{\langle ij \rangle \tau \tau'} \frac{1}{2} C_{ij}^\tau C_{ij}^{\tau'} K_{ij}   
\nonumber \\
&\times 
\left[{\cal N}(B^{\tau {\bar \tau} \dagger}_{ij} B^{\tau' \tau'}_{ij}) - A^{\tau {\bar \tau} \dagger}_{ij} A^{\tau' \tau'}_{ij} + H.c. \right], 
\label{eq:sb_si}
\end{align}
where ${\bar \tau}=\uparrow (\downarrow)$ for $\tau=\downarrow (\uparrow)$. 
The first, second, and third terms represent the intra-dimer electron hopping, the Heisenberg exchange interaction, and the spin-charge coupling, respectively. 
We define numerical factors as $C_{ij}^{\tau} = 1$ on the horizontal bonds along the $x$ axis, and $C_{ij}^{\tau} =1$ $(-1)$ for $\tau =\uparrow$ $ ( \downarrow)$ on other bonds. 

The MF decouplings are introduced in the quartic terms of the SB operators as  
$X_{ij} Y_{ij} \approx \langle X_{ij} \rangle Y_{ij} + X_{ij} \langle Y_{ij} \rangle - \langle X_{ij} \rangle \langle Y_{ij} \rangle$, where $X_{ij}$ and $Y_{ij}$ are $A_{ij}$ or $B_{ij}$. 
The local constraint is replaced by the global one given as 
$\sum_{i \sigma \tau} \beta^\dagger_{i \sigma \tau} \beta_{i \sigma \tau} = 2N$, where a summation with respect to $i$ is taken over on the A and B sublattices, $N$ is the number of the unit cells, and $\lambda$ is treated as the Lagrange multiplier. 
The MFs are assumed to be independent in the crystallographically inequivalent bonds labeled by $1$--$6$ in Fig.~\ref{fig:si}. 
We introduce a set of the boson operators defined in the momentum space as 
\begin{align}
\phi_{\bm k} &= (a_{{\bm k}\uparrow\uparrow}, a_{{\bm k}\downarrow\uparrow}, a_{{\bm k}\uparrow\downarrow}, a_{{\bm k}\downarrow\downarrow}, b_{{\bm k}\uparrow\uparrow}, b_{{\bm k}\downarrow\uparrow}, b_{{\bm k}\uparrow\downarrow}, b_{{\bm k}\downarrow\downarrow}, 
\nonumber \\
& 
a^{\dagger}_{-{\bm k}\uparrow\uparrow}, a^{\dagger}_{-{\bm k}\downarrow\uparrow}, a^{\dagger}_{-{\bm k}\uparrow\downarrow}, a^{\dagger}_{-{\bm k}\downarrow\downarrow}, b^{\dagger}_{-{\bm k}\uparrow\uparrow}, b^{\dagger}_{-{\bm k}\downarrow\uparrow}, b^{\dagger}_{-{\bm k}\uparrow\downarrow}, b^{\dagger}_{-{\bm k}\downarrow\downarrow}), 
\end{align}
where $a_{{\bm k} \sigma \tau}$ and $b_{{\bm k} \sigma \tau}$ are the SB operators for the A and B sublattices, respectively, in the momentum space. 
Then, the Hamiltonian in the SB MF approximation is expressed as 
\begin{align}
{\cal H}_{\rm SB} 
= \frac{1}{2} \sum_{\bm k} {\bm \phi}^{\dagger}_{\bm k} h_{\bm k} {\bm \phi}_{\bm k} - 6 \lambda N, 
\end{align}
where $h_{\bm k}$ is a $16 \times 16$ matrix. 
Through the Bogoliubov transformation, the Hamiltonian is diagonalized as 
\begin{align}
{\cal H}_{\rm SB} 
= \sum_{\bm k} \sum_{n=1}^{8} \omega_{{\bm k}n} 
\left (\eta^{\dagger}_{{\bm k} n} \eta_{{\bm k} n} + \frac{1}{2}\right ) - 6 \lambda N, 
\end{align}
with the energy $\omega_{{\bm k}n}$ and a set of the boson operators defined by the Bogoliubov transformation as $\eta_{{\bm k}}={\cal F}_{{\bm k}} \phi_{\bm k}$. 

The MFs are determined by the saddle-point equations given by  
\begin{align}
A_{n}^{\tau \tau'} 
&= \frac{1}{2N} \sum_{\bm k} e^{i{\bm k} \cdot {\bm \delta}} \sum_{i=1}^{8} 
\bigl[
({\cal F}^{-1}_{\bm k})_{m_{\uparrow\tau}+l_1,i}
({\cal F}^{-1}_{\bm k})^{\star}_{m_{\downarrow\tau'}+l_2,i} 
\nonumber \\
&- 
({\cal F}^{-1}_{\bm k})_{m_{\downarrow\tau}+l_1,i}
({\cal F}^{-1}_{\bm k})^{*}_{m_{\uparrow\tau'}+l_2,i} \bigr], 
\label{eq:AA}
\end{align}
and 
\begin{align}
B_{n}^{\tau \tau'} 
&= \frac{1}{2N} \sum_{\bm k} e^{-i{\bm k} \cdot {\bm \delta}}
\sum_{i=9}^{16} 
\bigl[
({\cal F}^{-1}_{\bm k})^{*}_{m_{\uparrow\tau}+h_1,i}
 ({\cal F}^{-1}_{\bm k})_{m_{\uparrow\tau'}+h_2,i} 
\nonumber \\
&+ 
({\cal F}^{-1}_{\bm k})^{*}_{m_{\downarrow\tau}+h_1,i}
({\cal F}^{-1}_{\bm k})_{m_{\downarrow\tau'}+h_2,i} \bigr], 
\label{eq:BB}
\end{align}
where 
$A^{\tau\tau'}_{n} \equiv \langle A^{\tau\tau'}_{ij} \rangle$ and 
$B^{\tau\tau'}_{n} \equiv \langle B^{\tau\tau'}_{ij} \rangle$, in which  
a bond connecting $i$ and $j$ sites belongs to the bond $n(=1$--$6)$ defined in Fig.~\ref{fig:si}. 
We define 
$(l_1. l_2)=(0, 8)$, $(4, 12)$, $(0, 12)$, $(4, 8)$, $(4, 8)$, and 
$(0, 12)$ in $A^{\tau\tau'}_{n}$ for $n=1$--$6$, 
and 
$(h_1. h_2)=(0, 0)$, $(4, 4)$, $(0, 4)$, $(4, 0)$, $(4, 0)$, and 
$(0, 4)$ in $B^{\tau\tau'}_{n}$ for $n=1$--$6$, respectively. 
We also define 
${\bm \delta}={\bm a}_{{\rm 1}+{\rm 2}}$, ${\bm a}_{{\rm 1}+{\rm 2}}$, ${\bm a}_{{\rm 1}}$, ${\bm a}_{{\rm 1}}$, ${\bm a}_{{\rm 2}}$ and ${\bm a}_{{\rm 2}}$ for $n=1$--$6$, respectively, 
where ${\bm a}_1$ and ${\bm a}_2$ are the primitive translation vectors given in Fig.~\ref{fig:si}, 
and  
$(m_{\uparrow \uparrow}, m_{\downarrow \uparrow}, m_{\uparrow \downarrow}, m_{\downarrow \downarrow}) = (1, 2, 3, 4)$.
The constraints in the two sublattices provide the following two equations:  
\begin{align}
\frac{1}{N}\sum_{\bm k} \sum_{i=1}^4\sum_{j=9}^{16} 
\left| ({\cal F}^{-1}_{\bm k})^{*}_{i,j} \right|^2
= 1, 
\label{eq:c1}
\end{align}
\begin{align}
\frac{1}{N}\sum_{\bm k} \sum_{i=5}^8\sum_{j=9}^{16} 
\left| ({\cal F}^{-1}_{\bm k})^{*}_{i,j} \right|^2  = 1 , 
\label{eq:c2}
\end{align}
through which the bond order parameters and the Lagrange multiplier are determined selfconsistently. 

From now on, we focus on the magnetic ordered phases. 
In the SB scheme, a long-range order at a wave vector $\bm Q$ 
corresponds to the Bose-Einstein condensation (BEC) for the boson $\alpha_{\bm k=\pm {\bm Q}/2}$~\cite{Hirsch_si}. 
When the BEC occurs by changing the parameter from the disordered state, 
the lowest excitation energy for the bosons at $\pm {\bm Q}/2$ is of the order of $1/N$. 
By following the conventional calculation process of the BEC, 
we separate the divergent terms at $\pm {\bm Q}/2$ from the summations for $\bm k$ 
in Eqs.~(\ref{eq:AA})--(\ref{eq:c2}) and solve the equations selfconsistently. 
The sublattice magnetizations are obtained from the BEC density at the momentum $\pm {\bm Q}$ given by 
\begin{align}
\rho_{\rm C} = \frac{1}{N} 
\sum_{{\bm k} = \pm {\bm Q}/2} 
\sum_{i=l}^{m} \sum_{j=9}^{16}  
\left|({\cal F}^{-1}_{\bm k})_{i,j } \right|^2  , 
\end{align}
where $(l, m)=(1,8)$ and $(9,16)$ for ${\rm C} = {\rm A}$ and $\rm B$, respectively. 
A condition $\rho_{\rm A} = \rho_{\rm B}$ is satisfied due to the equivalence of the A and B sublattices as mentioned in the previous section. 

\end{document}